\input harvmac.tex


\lref\Romans{L.~J.~Romans, ``Massive N=2A Supergravity in Ten
Dimensions", Phys.Lett. {\bf 169B} (1986) 374.}
\lref\PS{J.~Polchinski and A.~Strominger,``New Vacua for Type II
String Theory,'' Phys. Lett. {\bf B388} (1996) 736.}
\lref\deight{E.~Bergshoeff, M.~de Roo, M.~B.~Green, G.~Papadopoulos,
P.~K.~Townsend, ``Duality of Type II 7-branes and 8-branes",
Nucl.Phys. {\bf B470} (1996) 113.}

\lref\BGV{N.~Berkovits, S.~Gukov, and B.~Vallilo, to appear.}
\lref\GGWappear{S.J.~Gates, Jr., S.~Gukov and E.~Witten,
``Two Two-Dimensional Supergravity Theories from Calabi-Yau Four-Folds",
Nucl.Phys. {\bf B584} (2000) 109.}
\lref\HP{P.S.~Howe and G.~Papadopoulos, ``N=2, D = 2 Supergeometry",
Class. Quantum Grav. {\bf4} (1987) 11.}
\lref\GW{M.T.~Grisaru and M.E.~Wehlau, ``Prepotentials for (2,2)
Supergravity", Int. J. Mod. Phys. {\bf A10} (1995) 753.}
\lref\GGW{S.J.~Gates, Jr., M.T.~Grisaru and M.E.~Wehlau,
``A Study of General 2D, N=2 Matter Coupled to Supergravity in
Superspace", Nucl.Phys. {\bf B460} (1996) 579.}
\lref\GHR{S.J.~Gates, C.M.~Hull and M.~Rocek,
``Twisted Multiplets and New Supersymmetric Nonlinear Sigma Models",
Nucl.Phys. {\bf B248} (1984) 157.}
\lref\WB{E.~Witten and J.~Bagger,
``Quantization Of Newton's Constant In Certain Supergravity Theories",
Phys. Lett. {\bf 115B} (1982) 202.}
\lref\CJSFGN{E.~Cremmer, B.~Julia, J.~Scherk, S.~Ferrara, L.~Girardello and P.~van Nieuwenhuizen,
``Spontaneous Symmetry Breaking And Higgs Effect In Supergravity Without Cosmological Constant",
Nucl. Phys. {\bf B147} (1979) 105.}


\lref\GSW{M.~Green, J.~H.~Schwarz and E.~Witten, ``Superstring Theory",
Vol. 2, Cambridge Univ. Press, Cambridge 1988.}
\lref\GHM{M.~Green, J.H.~Harvey and G.~Moore, ``I-Brane Inflow
and Anomalous Couplings on D-branes",
Class. Quant. Grav. {\bf 14} (1997) 47.}
\lref\SV{A.~Strominger, C.~Vafa, ``Microscopic Origin of
the Bekenstein-Hawking Entropy", Phys.Lett. {\bf B379} (1996) 99.}
\lref\MSW{J.~Maldacena, A.~Strominger, E.~Witten,
``Black Hole Entropy in M-Theory", JHEP {\bf 9712} (1997) 002.}
\lref\Ventropy{C.~Vafa, ``Black Holes and Calabi-Yau Threefolds",
Adv.Theor.Math.Phys. {\bf 2} (1998) 207.}
\lref\Strominger{A.~Strominger, ``Black Hole Entropy from
Near-Horizon Microstates", JHEP {\bf 9802} (1998) 009.}
\lref\BTZ{M.~Banados, M.~Henneaux, C.~Teitelboim, J.~Zanelli,
``The Black Hole in Three Dimensional Space Time",
Phys.Rev.Lett. {\bf 69} (1992) 1849.}
\lref\Carlip{S.~Carlip, ``The (2+1)-Dimensional Black Hole",
Class.Quant.Grav. {\bf 12} (1995) 2853.}

\lref\AHKSS{O.~Aharony, A.~Hanany, K.~Intriligator, N.~Seiberg,
M.~J.~Strassler Aspects of N=2 Supersymmetric Gauge Theories in
Three Dimensions", Nucl.Phys. {\bf B499} (1997) 67.}
\lref\BHO{J.~de~Boer, K.~Hori, Y.~Oz, ``Dynamics of N=2
Supersymmetric Gauge Theories in Three Dimensions",
Nucl.Phys. {\bf B500} (1997) 163.}


\lref\Windex{E.~Witten, ``Constraints on Supersymmetry Breaking",
Nucl.Phys. {\bf B202} (1982) 253.}
\lref\Witten{E.~Witten, ``Non-Perturbative Superpotentials In
String Theory", Nucl.Phys. {\bf B474} (1996) 343.}
\lref\Wflux{E. Witten, ``On Flux Quantization In M-Theory And
The Effective Action", J. Geom. Phys. {\bf 22} (1997) 1.}
\lref\GVW{S.~Gukov, C.~Vafa and E.~Witten, ``CFT's From
Calabi-Yau Four-folds", hep-th/9906070.}
\lref\BB{K.~Becker, M.~Becker, ``M-Theory on Eight-Manifolds",
Nucl.Phys. {\bf B477} (1996) 155.}
\lref\SVW{S. Sethi, C. Vafa, and E. Witten,
``Constraints on Low-Dimensional String Compacti\-fi\-cations",
Nucl. Phys. {\bf B480} (1996) 213.}
\lref\DRS{K.~Dasgupta, G.~Rajesh, S.~Sethi,
``M Theory, Orientifolds and G-Flux", hep-th/9908088.}
\lref\HLouis{M.~Haack, J.~Louis, ``Aspects of Heterotic/M-Theory
Duality in D=3", hep-th/9908067.}
\lref\HM{J.A.~Harvey and G.~Moore,
``Superpotentials and Membrane Instantons", hep-th/9907026.}
\lref\Lerche{W.~Lerche, ``Fayet-Iliopoulos Potentials from
Four-Folds", JHEP {\bf 9711} (1997) 004.}
\lref\Mayr{P.~Mayr, ``Mirror Symmetry, N=1 Superpotentials and
Tensionless Strings on Calabi-Yau Four-Folds",
Nucl.Phys. {\bf B494} (1997) 489.}
\lref\Klemm{A.~Klemm, B.~Lian, S.-S.~Roan, S.-T.~Yau,
``Calabi-Yau fourfolds for M- and F-Theory compactifications",
Nucl.Phys. {\bf B518} (1998) 515.}


\lref\OOY{H.~Ooguri, Y.~Oz, Z.~Yin, ``D-Branes on Calabi-Yau
Spaces and Their Mirrors", Nucl.Phys. {\bf B477} (1996) 407.}
\lref\LVW{W.~Lerche, C.~Vafa and N.P.~Warner,
``Chiral Rings In N=2 Superconformal Theories",
Nucl. Phys. {\bf B324} (1989) 427.}
\lref\GMP{B.R.~Greene, D.R.~Morrison, M.R.~Plesser,
``Mirror Manifolds in Higher Dimension",
Commun.Math.Phys. {\bf 173} (1995) 559.}
\lref\Hubsch{T.~H\"ubsch, ``Some Algebraic Symmetries of
(2,2)-Supersymmetric Systems", hep-th/9903114.}
\lref\griff{P. Griffiths and J. Harris, {\it Principles Of
Algebraic Geometry} (Wiley-Interscience, 1978).}
\lref\Ktorus{M.~Kontsevich, ``Enumeration of rational curves
via torus actions", hep-th/9405035.}
\lref\Khom{M.~Kontsevich, ``Homological Algebra of
Mirror Symmetry", alg-geom/9411018.}
\lref\SYZ{A.~Strominger, S.-T.~Yau, E.~Zaslow,
``Mirror Symmetry is T-Duality", Nucl.Phys. {\bf B479} (1996) 243.}
\lref\Vafa{C.~Vafa, ``Extending Mirror Conjecture to Calabi-Yau
with Bundles", hep-th/9804131.}
\lref\KM{M.~Kontsevich and Yu.~Manin,
``Gromov-Witten classes, quantum cohomology, and enumerative geometry",
Commun.Math.Phys. {\bf 164} (1994) 525.}
\lref\Ginv{M.~Gromov, ``Pseudoholomorphic curves in symplectic manifolds",
Inv.Math. {\bf 82} (1985) 307.}
\lref\Winv{E.~Witten, ``Two-dimensional gravity and intersection
theory on moduli space", Surveys in Diff. Geom. {\bf 1} (1991) 243.}
\lref\GV{R.~Gopakumar, C.~Vafa, ``M-Theory and Topological
Strings -- I, II", hep-th/9809187, hep-th/9812127.}
\lref\Strom{A.~Strominger, ``Special geometry",
Comm. Math. Phys. {\bf 133} (1990) 163.}
\lref\Tian{G.~Tian, ``Smoothness of the universal deformation space
of compact Calabi-Yau manifolds and its Petersson-Weil metric",
in {\it Mathematical aspects of string theory},
World Sci. Publishing, Singapore, 1987, p. 629.}


\nref\HL{R.~Harvey and H.B.~Lawson, Jr., ``Calibrated geometries",
Acta Math. {\bf 148} (1982) 47.}
\nref\BBMOOY{K.~Becker, M.~Becker, D.R.~Morrison, H.~Ooguri, Y.~Oz, Z.~Yin,
``Supersymmetric Cycles in Exceptional Holonomy Manifolds and
Calabi-Yau 4-Folds", Nucl.Phys. {\bf B480} (1996) 225.}
\nref\GPcal{G.W.~Gibbons and G.~Papadopoulos, ``Calibrations and
Intersecting Branes", Commun. Math. Phys. {\bf 202} (1999) 593.}
\nref\GLWcal{J.P.~Gauntlett, N.D.~Lambert, P.C.~West, ``Branes and
Calibrated Geometries",  Commun.Math.Phys. {\bf 202} (1999) 571.}
\nref\AFFS{B.~S.~Acharya, J.~M.~Figueroa-O'Farrill, B.~Spence,
``Branes at angles and calibrated geometry", JHEP {\bf 9804} (1998) 012.}
\nref\FFcal{J.M.~Figueroa-O'Farrill,
``Intersecting brane geometries", hep-th/9806040.}
\nref\PTcal{G.~Papadopoulos and A.~Teschendorff, ``Grassmannians,
Calibrations and Five-Brane Intersections", hep-th/9811034.}
\nref\GPT{J.~Gutowski, G.~Papadopoulos, P.~K.~Townsend,
``Supersymmetry and Generalized Calibrations", hep-th/9905156.}
\nref\Minasian{R.~Minasian, D.~Tsimpis, ``M5-branes, Special
Lagrangian Submanifolds and sigma-models", hep-th/9906190.}

\lref\McLean{R.~C.~McLean, ``Deformations of Calibrated Submanifolds",
Comm. Anal. Geom. {\bf 6} (1998) 705.}
\lref\Hitchin{N.~Hitchin, ``The Moduli Space of Special Lagrangian
Submanifolds", Ann. Scuola Norm. Sup. Pisa Cl. Sci. (4) {\bf 25} (1997), 503.}
\lref\J{D.~Joyce, ``On counting special Lagrangian homology 3-spheres",
hep-th/9907013.}
\lref\Gross{M.~Gross, ``Special Lagrangian Fibrations I: Topology",
alg-geom/9710006.}
\lref\Acharya{B.~S.~Acharya, ``M theory, Joyce Orbifolds and Super
Yang-Mills", hep-th/9812205.}


\let\includefigures=\iftrue
\newfam\black
\includefigures
\input epsf
\def\figin{\epsfcheck\figin}\def\figins{\epsfcheck\figins}
\def\epsfcheck{\ifx\epsfbox\UnDeFiNeD
\message{(NO epsf.tex, FIGURES WILL BE IGNORED)}
\gdef\figin##1{\vskip2in}\gdef\figins##1{\hskip.5in}
\else\message{(FIGURES WILL BE INCLUDED)}%
\gdef\figin##1{##1}\gdef\figins##1{##1}\fi}
\def\DefWarn#1{}
\def\figinsert{\goodbreak\midinsert}
\def\ifig#1#2#3{\DefWarn#1\xdef#1{fig.~\the\figno}
\writedef{#1\leftbracket fig.\noexpand~\the\figno}%
\figinsert\figin{\centerline{#3}}\medskip\centerline{\vbox{\baselineskip12pt
\advance\hsize by -1truein\noindent\footnotefont{\bf Fig.~\the\figno:} #2}}
\bigskip\endinsert\global\advance\figno by1}
\else
\def\ifig#1#2#3{\xdef#1{fig.~\the\figno}
\writedef{#1\leftbracket fig.\noexpand~\the\figno}%
\global\advance\figno by1}
\fi

\font\cmss=cmss10 \font\cmsss=cmss10 at 7pt

\def\IB{\relax\hbox{$\inbar\kern-.3em{\rm B}$}}
\def\IC{\relax\hbox{$\inbar\kern-.3em{\rm C}$}}
\def\ID{\relax\hbox{$\inbar\kern-.3em{\rm D}$}}
\def\IE{\relax\hbox{$\inbar\kern-.3em{\rm E}$}}
\def\IF{\relax\hbox{$\inbar\kern-.3em{\rm F}$}}
\def\IG{\relax\hbox{$\inbar\kern-.3em{\rm G}$}}
\def\IGa{\relax\hbox{${\rm I}\kern-.18em\Gamma$}}
\def\IH{\relax{\rm I\kern-.18em H}}
\def\IK{\relax{\rm I\kern-.18em K}}
\def\IL{\relax{\rm I\kern-.18em L}}
\def\IP{\relax{\rm I\kern-.18em P}}
\def\IR{\relax{\rm I\kern-.18em R}}
\def\Z{\relax\ifmmode\mathchoice
{\hbox{\cmss Z\kern-.4em Z}}{\hbox{\cmss Z\kern-.4em Z}}
{\lower.9pt\hbox{\cmsss Z\kern-.4em Z}}
{\lower1.2pt\hbox{\cmsss Z\kern-.4em Z}}\else{\cmss Z\kern-.4em
Z}\fi}

\def\II{\relax{\rm I\kern-.18em I}}

\def\S{{\bf S}}


\def\CF {{\cal F}}

\def\CI {{\cal I}}

\def\CK {{\cal K}}

\def\CM {{\cal M}}
\def\CN {{\cal N}}

\def\CV {{\cal V}}

\def\CX {{\cal X}}


\def\p{\partial}

\def\tilde{\widetilde}
\def\hat{\widehat}
\def\bar{\overline}


\def\Tr{{\rm Tr}}

\def\p{\partial}

\def\inbar{\,\vrule height1.5ex width.4pt depth0pt}
\def\r{{\rm Re}}

\def\a{\alpha}

\def\th{\theta}
\def\s{\sigma}

\def\p{\partial}
\def\S{\Sigma}


\Title{ \vbox{\baselineskip12pt
\hbox{PUPT-1881}
\hbox{ITEP-TH-38/99}
\hbox{CALT-68-2244}
\hbox{CITUSC/005}}}
{{\vbox{\centerline{Solitons, Superpotentials and Calibrations}}}}
\centerline{Sergei Gukov\foot{On leave from the Institute of
Theoretical and Experimental Physics and the L.~D.~Landau
Institute for Theoretical Physics}}
\vskip 10pt
\centerline{\it Joseph Henry Laboratories, Princeton University}
\centerline{\it Princeton, New Jersey 08544}
\centerline{\it and}
\centerline{\it California Institute of Technology, Pasadena, California 91125}
\centerline{\it CIT-USC Center For Theoretical Physics}
\vskip 15pt
{\bf \centerline{Abstract}}

In this paper we study several issues related to the generation
of superpotential induced by background Ramond-Ramond fluxes
in compactification of Type IIA string theory on Calabi-Yau four-folds.
Identifying BPS solitons with D-branes wrapped over calibrated submanifolds
in a Calabi-Yau space, we propose a general formula for the superpotential
and justify it comparing the supersymmetry conditions in $D=2$
and $D=10$ supergravity theories.
We also suggest a geometric interpretation to the supersymmetric index
in the two-dimensional effective theory in terms of topological invariants
of the Calabi-Yau four-fold, and estimate the asymptotic growth of these
invariants from BTZ black hole entropy.
Finally, we explicitly construct new supersymmetric vacua
for Type IIA string theory compactification on a Calabi-Yau
four-fold with Ramond-Ramond fluxes.

\Date{November 1999}

\newsec{Introduction and Summary}

The theory of calibrations was developed by Harvey and Lawson \HL\
for the purpose of obtaining geometries with the property that
all varieties in the geometry are volume minimizing.
Due to this fundamental property, calibrations have proven to be
a very useful tool for investigating supersymmetric configurations
of D-branes in string theory.
A partial list of references includes \refs{\BBMOOY - \Minasian}.
Although a significant progress has been made in the study
of the local differential geometry of the moduli space of
calibrated submanifolds \refs{\McLean,\Hitchin},
its global structure remains elusive.
Motivated by \refs{\SYZ,\Vafa},
we hope that physical applications of calibrated geometry
can help to address this problem.

In this paper we study compactification of Type IIA string theory
on a Calabi-Yau four-fold with background Ramond-Ramond fluxes.
In particular, we are interested in the vacuum and soliton
structure of the resulting $\CN=(2,2)$ theory in two dimensions.
We show that supersymmetric vacua can be described by
the effective superpotential generated by Ramond-Ramond fields,
while the BPS solitons connecting these vacua correspond to D-branes
wrapped over supersymmetric cycles in the Calabi-Yau space $X$.
Using the relation between supersymmetric cycles and
calibrations we derive the general formula for the superpotential
in the effective field theory:
$$
W = \sum \int_X ({\rm calibrations}) \wedge ({\rm Ramond-Ramond~ fields})
$$
There are two basic types of calibrations on a Calabi-Yau four-fold $X$
given by the covariantly constant forms $\r (\Omega)$ and
${1 \over p!} \CK^p$, where $\Omega$ denotes the nowhere-vanishing
holomorphic $(4,0)$-form and $\CK$ is the K\"ahler form on $X$.
The first type of calibrations corresponds to special Lagrangian
submanifolds in $X$ and leads to a chiral superpotential,
while the latter is associated with holomorphic cycles of complex
dimension $p$ and generates a twisted chiral superpotential.

Given the induced superpotential $W$ one can define
a supersymmetric index $\CI_{[S]} (X)$ that counts the ``number"
of supersymmetric cycles with the homology class $[S] \in H_* (X)$.
Namely, $\CI_{[S]} (X)$ is equal to the Euler number of the moduli
space of calibrated submanifolds $S$ with flat line bundles:
$$
\CI_{[S]} (X) = \chi \Big( \CM_{[S]} (X) \Big)
$$
We argue that $\CI_{[S]} (X)$ is a topological invariant of
the Calabi-Yau four-fold $X$.
The dimension of $X$ is very important here, so we do not expect
$\CI_{[S]} (X)$ to generalize directly to higher (or lower) dimensional
Calabi-Yau manifolds (however, see \J\ and \GV\ where analogous
invariants of Calabi-Yau three-folds have been discussed).
In some special cases, $\CI_{[S]} (X)$ can be shown to agree
with classical topological invariants of $X$.
For example, when $[S]$ is the class of a special Lagrangian
torus, from the conjecture of Strominger, Yau and Zaslow \SYZ\
it follows that $\CI_{[S]} (X) = \chi (X)$.

The paper is organized as follows.
In the next section we describe the relevant aspects of $\CN=(2,2)$
theories constructed from Calabi-Yau four-folds and derive
the effective superpotential generated by Ramond-Ramond fluxes.
In section 3 we identify BPS solitons in the effective theory
with D-branes wrapped over calibrated submanifolds in Calabi-Yau spaces.
We also discuss possible generalizations and conjecture the form
of the superpotential generated by background fields in
compactification on manifolds with exceptional holonomy groups,
$G_2$ and $Spin(7)$.
Superpotentials induced by membrane instantons in
$G_2$-manifolds have been discussed in \Acharya\ and \HM.
Section 4 is devoted to the geometrical interpretation of
the supersymmetric index $\CI_{[S]} (X)$.
In the case when $S$ is a holomorphic curve, we estimate
the asymptotic growth of $\CI_{[S]} (X)$ from the counting
of BTZ black hole microstates.
In section 5 we explicitly construct new supersymmetric
vacua for Type IIA string theory compactification on
Calabi-Yau four-folds with Ramond-Ramond fluxes.
A work along these lines has been presented recently in \DRS.
Finally, in the appendix we justify the formula for the induced
superpotential comparing the supersymmetry conditions
in $D=2$ and $D=10$ supergravity theories.

Throughout the paper we assume that $X$ is a smooth compact
Calabi-Yau four-fold, so that all the vacua in the effective
two-dimensional theory are connected by solitons.
Most of the results presented here can be generalized to
a non-compact $X$, though not all the vacua are connected
by solitons in such models \GVW.


\newsec{Superpotentials from Calabi-Yau Four-folds}

Compactification of Type IIA string theory on a Calabi-Yau four-fold
$X$ leads to $\CN=(2,2)$ theory in two dimensions.
The low-energy spectrum of this theory includes $h^{3,1}$ chiral
superfields $\Phi_i$ and $h^{1,1}$ twisted chiral superfields $\S_j$.
Here we use the standard notation $h^{p,q}$ for
the dimension of the Hodge group $H^{p,q}_{\bar \p} (X)$.
We denote by $\phi_i$ and $\s_j$ the scalar components of the superfields
$\Phi_i$ and $\S_j$ which correspond, respectively, to deformations of
the complex and the K\"ahler structure of $X$.
Note, that the mirror symmetry maps Type IIA string theory on
a four-fold $X$ also to Type IIA string theory on the mirror
variety $\tilde X$, such that:
\eqn\hmirror{ h^{p,q} (X) = h^{4-p,q} (\tilde X)}
and conformal field theories associated with $X$ and $\tilde X$ are equivalent.
In the two-dimensional effective theory this operation corresponds
to the twist that exchanges the multiplets $\Phi_i$ and $\S_j$.

Let us assume for a moment that there are no background fluxes.
Then the two-dimensional effective theory is described by
$\CN=(2,2)$ dilaton supergravity interacting with a non-linear sigma-model.
The complete superspace formulation of this theory will be presented
elsewhere \GGWappear.
The target space of this sigma-model is parametrized by
the chiral fields $\{ \phi_i \}$ and the twisted chiral fields $\{ \s_j \}$,
so that the target space metric is K\"ahler and torsionless.
{}From the Kaluza-Klein reduction of Type IIA theory on $X$
one can easily find that in the large volume limit
the target space metric is equal to the metric on
the moduli space of $X$, $\CM_c (X) \times \CM_{\CK} (X)$.
For example, the effective K\"ahler potential for the chiral superfields:
\eqn\kforphi{K_c (\Phi_i, \bar \Phi_{\bar i}) =
- \log \Big( \int_X \Omega \wedge \bar \Omega \Big),}
where $\Omega$ denotes the covariantly constant $(4,0)$-form, is equal to
the K\"ahler potential of the Weil-Petersson metric on $\CM_c (X)$.
Therefore, at least classically, we may identify the space
of vacua with the moduli space of the Calabi-Yau manifold $X$.
However, as we explain below, most of these vacua are lifted
once we turn on Ramond-Ramond fluxes.

One of the most important features of low-dimensional string theory
compactifications is the global anomaly \refs{\SVW,\Wflux}
given by the Euler number, $\chi (X)$, of the Calabi-Yau four-fold $X$.
In general, to cancel the anomaly one has to introduce
a background flux for the four-form field $G$ and/or
$N$ fundamental strings filling two-dimensional space-time,
such that the following condition is satisfied:
\eqn\sethrel{N={\chi\over 24}-{1\over 2 (2\pi)^2}\int_X G \wedge G}
In particular, $G$ must be non-zero when $\chi(X)/24$ is not integer.
Besides $G$-fluxes one can also turn on background values
of the other Ramond-Ramond fields: a 2-form $F$ and
a zero-form $M$ dual to the ten-form associated with D8-branes.
Note, non-zero value of $M$ leads to the ten-dimensional
theory with a cosmological constant \PS, the so-called
massive Type IIA supergravity \Romans.
Furthermore, one can consider compactifications where background
fields $F$ and $G$ have two indices in the noncompact directions.
In what follows it will be convenient, instead, to introduce
the dual eight-form $\check F^{(8)}$ and the six-form $\check G^{(6)}$
which have only internal space-time indices.
All these Ramond-Ramond field strengths can be combined into a formal sum:
$$
\CF = M + i F + G + i \check G^{(6)} + \check F^{(8)},
$$
so that $\CF \in H^* (X)$.
We denote by $\CV_k$ (resp. $\nu_k$) two-dimensional
superfields (resp. their scalar components)
representing different choices of the flux $\CF$.

In general, non-zero Ramond-Ramond fluxes break
supersymmetry generating an effective superpotential
$W (\Phi_i, \S_j, \CV_k)$ in the two-dimensional theory \refs{\GVW,\Lerche}.
Usually it is very difficult to see the superfields $\CV_k$ explicitly,
so we mainly consider the chiral superpotential $W (\Phi_i)$
and the twisted chiral superpotential $\tilde W (\S_j)$
obtained via integrating over the fields $\CV_k$.
A simple way to find effective superpotential
is to interpret BPS solitons in $\CN=(2,2)$ theory
as D-branes wrapped over supersymmetric cycles in $X$.

Let us start with a simple example which is
actually a precursor of calibrated geometry.
Consider compactification of Type IIA theory
on $X$ with a non-zero flux of the 4-form field strength $G$.
The effective superpotential looks like \GVW:
\eqn\wone{ W(\Phi_i) = {1 \over 2 \pi} \int_X \Omega \wedge G}
To see how the formula \wone\ comes about, let us take a D4-brane
wrapped over a supersymmetric four-cycle $S \in H_4 (X, \Z)$.
In the two-dimensional effective field theory this state is a
BPS soliton interpolating between two supersymmetric vacua.
Since $G$ jumps across the D4-brane, these vacua correspond
to the different four-form fluxes $G_1$ and $G_2$, such that
$\Delta G / 2 \pi= (G_1 - G_2)/ 2 \pi$
is Poincar\'e dual to the homology class $[S]$.
In order to find the superpotential we note that in the effective
$\CN=(2,2)$ theory the mass of the BPS soliton connecting
the two vacua is given by the absolute value of $\Delta W$.
On the other hand, the mass of this soliton is given by the mass
of the D4-brane wrapped over the special Lagrangian cycle $S$.
Therefore we find $\Delta W = \int_S \Omega = {1 \over 2 \pi}
\int_X \Omega \wedge \Delta G$.

Regarded as a chiral primary state in the Hilbert space of
the $\CN=2$ SCFT associated with the Calabi-Yau space $X$,
$\Delta W = \int_S \Omega$ has the mirror non-chiral counterpart \OOY:
\eqn\dwtwo{\Delta \tilde W (\S_j) = {1 \over 2 \pi}
\int_{\tilde X} e^{\tilde \CK} \wedge \Delta \CF}
where $\tilde \CK$ is a complexified K\"ahler class
of the mirror Calabi-Yau manifold $\tilde X$.
In the dual two-dimensional effective theory obtained by
compactification on $\tilde X$, it is natural to interpret
\dwtwo\ as a change of twisted superpotential.
Therefore, by mirror symmetry we expect the following
twisted chiral superpotential:
\eqn\wtwo{\tilde W (\S_j) = {1 \over 2 \pi} \int_X e^{\CK} \wedge \CF}
in Type IIA compactification on a Calabi-Yau four-fold $X$.

There are several terms in \wtwo\ which also can be
deduced from the relation between solitons and D-branes.
Consider a soliton constructed from a D8-brane
wrapped over the entire $X$. Since the value of $M$ changes by
$2 \pi$ in crossing the D8-brane, the soliton connects two vacua
with different values of the field $M$.
In the theory with four supercharges the mass of this
soliton ${\rm ~ Vol}(X) = {1 \over 4!} \int_X \CK \wedge
\CK \wedge \CK \wedge \CK$ is expected to be proportional
to the change in the superpotential $\Delta \tilde W$.
This is indeed the case, in accordance with the
formula \wtwo\ predicted by the mirror symmetry.
In order to obtain another term in \wtwo\ proportional to the eight-form
flux one has to consider a soliton constructed from a D0-brane.
Since the latter does not wrap any cycle in the Calabi-Yau space $X$,
the mass of this soliton does not depend on the moduli of $X$,
in accordance with $\Delta \tilde W = {1 \over 2 \pi} \int_X \check F^{(8)}$.

The induced superpotential lifts (part of) the classical vacua
in the effective two-dimensional theory.
Now a point in the moduli space $\CM_c (X) \times \CM_{\CK} (X)$
corresponds to a supersymmetric vacuum only if it is a minimum
of the effective superpotential.
This picture suggests to interpret the extremality conditions of the
induced superpotential as the supersymmetry conditions in Type IIA
theory on the Calabi-Yau space $X$ with Ramond-Ramond fluxes.
Let us illustrate this interpretation with the following example.
Since Type IIA string theory is related to M-theory via
compactification on a circle we expect Type IIA supersymmetry
conditions in part to be similar to those in
eleven-dimensions \refs{\GVW,\BB}.
For instance, if the 4-form $G$ is the only non-vanishing
Ramond-Ramond field the supersymmetry conditions should be
exactly the same as in M-theory.
When $G$ has only internal space-time indices,
eq. \wtwo\ leads to a simple expression for the superpotential
proposed in \GVW:
\eqn\kkg{ \tilde W = {1 \over 4\pi} \int_X \CK \wedge \CK \wedge G}
Its variation is equivalent to the primitivity condition
$G \wedge \CK = 0$ which is one of the supersymmetry
conditions found in \BB.
The other supersymmetry constraints require $G^{(3,1)}=0$.
This is precisely what one finds from the variation of
the superpotential \wone.

More generally, we expect the supersymmetry conditions in Type IIA
theory on $X$ with Ramond-Ramond fluxes to be equivalent to the
supersymmetry conditions in the two-dimensional effective field theory.
A supersymmetric vacuum of the two-dimensional theory corresponds
to a minimum of the superpotential:
\eqn\minimum{ {D W \over D \phi_i} = 0 \quad {\rm and}
\quad {D \tilde W \over D \sigma_i} = 0,}
where we use the appropriate covariant derivatives
$D_{\phi_i} = D/D \phi_i$ and $D_{\sigma_j} = D/D \sigma_j$.
The form of the covariant connection is determined by the requirement
that $DW$ transforms covariantly under a K\"ahler transformation
$K \to K + F + \bar F$, where $F$ is an arbitrary holomorphic function.
Since the superpotential transforms as $W \to e^{-F} W$,
we must define the holomorphic connection as
$DW = \p W + (\p K) W$ in order to have $DW \to e^{-F} DW$
under the K\"ahler transformation \WB.
In plain english, this means that in supergravity theory with four
supercharges a superpotential is a section of a holomorphic line bundle,
rather than a usual function.
Therefore, the chiral superpotential \wone\ proportional to the
choice of $\Omega$ (which is a section of a line bundle over $\CM_c (X)$)
exhibits exactly the right behavior.

In a supersymmetric vacuum
either $W$ or $\tilde W$ should vanish\foot{A simple way to see
this using techniques of the present section
is to consider a composite domain wall that interpolates between
this vacuum and a vacuum with no flux, $\CF=0$.
Since the values of both $W$ and $\tilde W$ should jump
across this domain wall, from the general relation between
fluxes and corresponding D-branes it follows that the composite
domain wall must be constructed from two kinds of D-branes ---
D-branes wrapped over special Lagrangian cycles, and
D-branes wrapped over holomorphic cycles.
Each of these D-branes is BPS since it is wrapped on a supersymmetric cycle.
However, the two kinds of D-branes preserve different supersymmetry,
so that in total they break supersymmetry completely.} and,
moreover, all the low-energy multiplets must
be massive in a vacuum with non-zero superpotential \GVW.
The value of $W$ (resp. $\tilde W$) in the minimum determines
the mass of the gravitino fields \refs{\GGWappear,\GGW}.
Therefore, in general we expect that compactifications of
massive Type IIA supergravity on a Calabi-Yau four-fold $X$
lead to two-dimensional vacua with negative cosmological constant given by:
\eqn\cconst{\Lambda = - \vert W \vert^2}
Actually, more careful analysis shows that these vacua typically have
non-constant dilaton field in the background \BGV.

One can directly check that ten-dimensional supersymmetry
conditions are equivalent to the equations \minimum\ and \cconst,
with the holomorphic functions $W$ and $\tilde W$ given by \wone\ and \wtwo.
For example, extremizing \wone\ and \wtwo\ in flat space
we find the following simple equations\foot{Notice,
the holomorphic covariant derivative
$D_{\phi_i} W  = \p_{\phi_i} W + (\p_{\phi_i} K_c (\phi_i)) W$
computed using the K\"ahler potential \kforphi\ has a nice property
that $D_{\phi_i} \Omega$ is orthogonal to $\Omega$ \refs{\Strom, \Tian}.
Therefore, one can equivalently define $D_{\phi_i}$ as a covariant connection
which satisfies $\int_X (D_{\phi_i} \Omega) \wedge \bar \Omega =0$.}:
\eqn\scone{G^{(3,1)} = G^{(1,3)}=0}
and
\eqn\sctwo{ {M \over 6} \CK \wedge \CK \wedge \CK +
{i\over 2} F \wedge \CK \wedge \CK + G \wedge \CK
+ i \check G^{(6)} = 0}
which are precisely the supersymmetry conditions that one
finds in compactification of massive Type IIA supergravity
on ${\bf R}^2 \times X$ with background Ramond-Ramond fluxes.
In a vacuum with non-zero cosmological constant supersymmetry
conditions are more subtle due to the dependence on K\"ahler potential.

Let us now see what kind of corrections we could
expect to the semi-classical expressions \wone\ and \wtwo.
Because the Type IIA dilaton comes in the real gravitational
multiplet there are no stringy corrections to the superpotential.
Furthermore, the chiral superpotential $W(\Phi_i)$ is exact at the tree
level simply because there are no three-branes in Type IIA string theory.
On the other hand, the twisted chiral superpotential $\tilde W(\S_j)$
is modified by world-sheet instantons \Lerche\
as well as by five-brane instantons \Witten.
Since a five-brane wrapped over a cycle $S \in H_6 (X,\Z)$ implies
the cohomology class of the 4-form field $G$ restricted to $S$ to be
trivial, non-zero 4-flux can prevent five-brane instanton corrections.
Nevertheless, one can use the mirror symmetry to find the exact
form of the twisted chiral superpotential \Lerche.
The basic idea is that
effective two-dimensional theories constructed by considering
Type IIA theory on the mirror Calabi-Yau four-folds $X$ and $\tilde X$
must be equivalent. In particular, we expect:
\eqn\wwmirror{W_{X} (\Phi_i) = \tilde W_{\tilde X} (\Sigma_i) }
Therefore, the world-sheet instanton effects (on the right-hand
side of the formula \wwmirror) can be found evaluating the period
integral \wone\ on the left-hand side of \wwmirror.


\newsec{Solitons and Calibrations}

In general, we can consider a $D(2p)$-brane of Type IIA theory
wrapped over a $2p$-cycle $S \in H_{2p} (X,\Z)$.
In the effective two-dimensional field theory this state looks like
a soliton connecting two vacua corresponding to the different set of
Ramond-Ramond fluxes on the eight-manifold $X$.
The difference between the fluxes is given by the Poincar\'e dual
cohomology class $\Delta \CF = \hat{[S]} \in H^{8-2p} (X,\Z)$.

Furthermore, BPS solitons correspond to D-branes wrapped over
supersymmetric cycles. Out of the states with a given set of charges,
a soliton that saturates the Bogomolnyi bound has the least mass.
In string theory it means that the corresponding D-brane is wrapped
over a cycle which has minimal area in its homology class.
This property of BPS states constructed from wrapped D-branes has
a very nice geometric interpretation in terms of calibrations \HL.

Let us remind the definition of calibrated submanifolds,
as in Harvey and Lawson \HL. Let $\Psi$ be a closed $k$-form on $X$.
We say that $\Psi$ is a calibration if it is less than or equal to
the volume on each oriented $k$-dimensional submanifold $S \subset X$.
Namely, combining the orientation of $S$ with the restriction of
the Riemann metric on $X$ to the subspace $S$, we can define
a natural volume form ${\rm vol} (T_x S)$ on the tangent
space $T_x S$ for each point $x \in S$.
Then, $\Psi \vert_{T_x S} = \a \cdot {\rm vol} (T_x S)$
for some $\a \in \IR$, and we write:
$$
\Psi \vert_{T_x S} \le {\rm vol} (T_x S)
$$
if $\a \le 1$. If equality holds for all points $x \in S$,
then $S$ is called a calibrated submanifold with respect
to the calibration $\Psi$. According to this definition, the volume
of a calibrated submanifold $S$ can be expressed in terms of $\Psi$ as:
\eqn\Svol{ {\rm Vol} (S) = \int_{x \in S} \Psi \vert_{T_ x S} =
\int_S \Psi = \int_X \Psi \wedge \hat{[S]} }
Since the right-hand side depends only on the cohomology
class, we can write:
$$
{\rm Vol} (S) = \int_X \Psi \wedge \hat{[S]} =
\int_X \Psi \wedge \hat{[S']} =
\int_{x \in S'} \Psi \vert_{T_ x S'}
\le \int_{x \in S'} {\rm vol} (T_x S') = {\rm Vol} (S')
$$
for any other submanifold $S'$ in the same homology class.
Therefore, we have just demonstrated that calibrated manifolds
have minimal area in their homology class, the property we
expect of the wrapped D-branes to represent BPS solitons.

{}From the supersymmetry algebra it follows that the mass of a BPS
soliton is equal to the change in superpotential, $\vert \Delta W \vert$.
Since the former is given by the formula \Svol\
and the change of the Ramond-Ramond fluxes $\Delta \CF$
is Poincar\'e dual to the homology class $[S]$,
we obtain the following general formula for the superpotential:
\eqn\calibr{ W (\Phi_i, \S_j) = {1 \over 2 \pi} \int_X \Psi \wedge \CF}

Calibrations are very useful in the study of manifolds with
special holonomy groups. This is because special holonomy
groups are characterized by the existence of covariantly
constant forms. These forms can be used as calibrations.
Let us now illustrate the formula \calibr\ with some examples
of special calibrations.

$\underline{SU(4)~ {\rm Calibrations}:}$
The main goal of this paper is to study Type IIA compactifications
on Calabi-Yau four-folds. These manifolds have $SU(4)$ holonomy group.
There are two types of calibrations on manifolds with
$SU(4)$ holonomy group (more generally, on manifolds with
$SU(n)$ holonomy  group). If we take:
\eqn\ccayley{ \Psi = {1 \over 2} \CK^2 + \r (e^{i \th} \Omega),}
we obtain a $U(1)$ family of the so-called Cayley calibrations.
Substituting the Cayley calibration
into \calibr, we reproduce the chiral
superpotential \wone\ plus the twisted chiral superpotential \kkg.
This is, of course, not surprising since deriving the formula
\wone\ we implicitly used the special case of the calibration
\ccayley\ corresponding to special Lagrangian submanifolds.
Indeed, a special Lagrangian manifold is a calibrated submanifold
with respect to the special Lagrangian calibration:
$$
\Psi = \r (\Omega)
$$

One can obtain the other type of $SU(4)$ calibrations ---
K\"ahler calibrations ---
considering various powers of the complexified K\"ahler form:
\eqn\ckahler{ \Psi = {1 \over p!} \CK^p }
Since $\CK$ is covariantly constant, \ckahler\ is covariantly
constant as well. The submanifolds calibrated by this $\Psi$
are complex submanifolds $S \subset X$ of complex dimension $p$.
It is easy to see that in this case the general formula
\calibr\ yields the twisted chiral superpotential \wtwo\
predicted by mirror symmetry, with the right numerical coefficients.

$\underline{G_2~ {\rm Calibrations}:}$
In our derivation of the superpotential \calibr\ we made
very mild assumptions about the geometry of the manifold $X$.
Namely, $X$ had to be a Riemannian manifold, not necessarily
eight-dimensional, such that a compactification of Type IIA
string on $X$ preserved four unbroken supersymmetries.
Hence, all of the above arguments also apply
verbatim to the seven-dimensional
manifolds of $G_2$ holonomy, with the word `a soliton' replaced
by `a domain wall'. This change is due to the fact that
now we talk about $\CN=2$ three-dimensional effective field theory,
so that different vacua are connected by domain walls rather than by solitons.

Manifolds of $G_2$ holonomy are characterized by a covariantly
constant three-form $\Psi^{(3)}$ invariant under the exceptional
group $G_2$. From the formula \calibr\ we expect to find the following
superpotential in the effective three-dimensional theory:
\eqn\gtwow{ W = {1 \over 2 \pi} \int_X \Psi^{(3)} \wedge G }
A domain wall that connects different vacua in this theory
corresponds to a $D4$-brane wrapped over three-dimensional
associative submanifold $S$ calibrated by $\Psi^{(3)}$.

The coassociative calibration $\Psi^{(4)} = \star \Psi^{(3)}$ is defined
as a 4-form Hodge dual to $\Psi^{(3)}$. The submanifolds calibrated
by $\Psi^{(4)}$ are four-dimensional coassociative submanifolds.
Since a domain wall in three non-compact dimensions can be
constructed from an NS5-brane wrapped over a coassociative
cycle $S$, from the expression \calibr\ we expect the following
superpotential corresponding to the calibration $\Psi^{(4)}$:
\eqn\gtwohw{W = {1 \over 2 \pi} \int_X \Psi^{(4)} \wedge H}
Here $H$ is the NS-NS three-form field strength.

$\underline{Spin(7)~ {\rm Calibrations}:}$
Even though our arguments do not directly apply to the case of
eight-manifolds with $Spin(7)$ holonomy which break too much
supersymmetry, it is amusing to employ the general formula
\calibr\ to the Cayley calibration $\Psi^{(+)}$ of degree four.
The form $\Psi^{(+)}$ is self-dual, $\Psi^{(+)} = \star \Psi^{(+)}$.
We conjecture the following expression for the superpotential:
\eqn\spinw{ W = {1 \over 2 \pi} \int_X \Psi^{(+)} \wedge G }
In the present paper we will not pursue the proof of
the formulas \gtwow\ - \spinw\ for the manifolds of the exceptional holonomy.

As one can definitely see, calibrated geometries prove to be very
useful in writing superpotentials induced by background fluxes.

\newsec{A Mathematical Application: Counting BPS Solitons}

To explain the geometric meaning of the effective superpotentials
found in the previous sections, we recall that the expectation
values of the fields $\phi_i$ and $\sigma_j$ play the role of the
K\"ahler and the complex moduli of a Calabi-Yau space $X$, respectively.
Therefore, in a background with zero Ramond-Ramond fields
the classical space of supersymmetric vacua of the two-dimensional
effective field theory is simply the moduli space of $X$.

When Ramond-Ramond fluxes do not vanish, the degeneracy of
supersymmetric ground states is (partly) lifted by
the induced superpotential
\foot{For the purposes of the present section we restore
dependence on the moduli $\nu_k$ of the Ramond-Ramond fields.
Since in many cases the explicit form of the complete
superpotential $W (\Phi_i, \S_j, \CV_k)$ is not known, we do not
differentiate between chiral and twisted chiral superfields.}
$W(\Phi_i, \S_j, \CV_k)$.
Solving the equations:
\eqn\minv{ {D W \over D \nu_k} = 0}
for $\nu_k$, one can integrate out the fields $\CV_k$ and obtain
the effective chiral superpotential $W (\Phi_i)$ and
the twisted chiral superpotential $\tilde W (\S_j)$.
If for certain values of $\phi_i$ and $\sigma_j$ the resulting
superpotentials have a minimum,
then the two-dimensional field theory has a supersymmetric vacuum
obtained from the supersymmetric compactification on $X$ with
the K\"ahler and the complex moduli defined by the solution of \minimum.
Thus, supersymmetric vacua form a subspace in the moduli
space of $X$, as shown in Fig.1.
For example, a compactification with a $G$-flux
is supersymmetric only if the K\"ahler structure of $X$ is
such that $G$ is primitive \BB. This condition is equivalent
to \minimum\ with the twisted superpotential \kkg.

\ifig\pic{Classical vacua of the effective two-dimensional
field theory form a subspace in the moduli space of $X$.
Every point of this subspace is associated with
the moduli space of a supersymmetric cycle $S$,
for a given complex and K\"ahler structure of $X$.}
{\epsfxsize4.0in\epsfbox{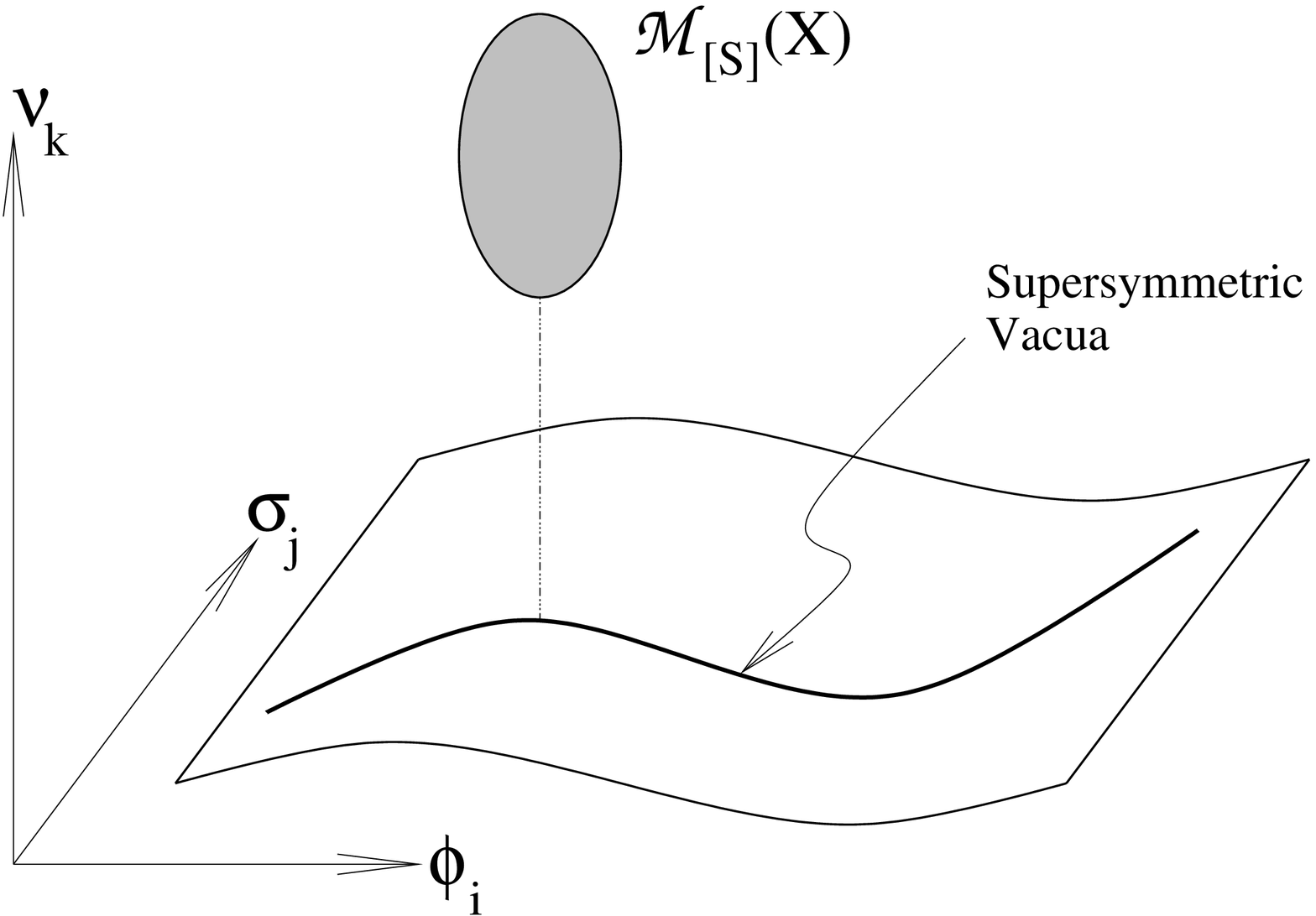}}

Notice that the superpotentials \wone\ and \wtwo\
constructed in the previous section depend only on
the cohomology class of the Ramond-Ramond flux $\CF$.
Therefore, when \minv\ has several solutions, a minimum
of the effective superpotential with the fields $\CV_k$
integrated out corresponds to multiple supersymmetric
vacua, all representing the same cohomology class $[\CF]$.
This multiplicity appears, for example, in a computation
of the Witten index in the effective two-dimensional theory.
For every point $(\phi_i, \sigma_j )$ in the moduli
space of $X$, let us denote its contribution to the supersymmetric
index by $\CI_{[\CF]} (X)$:
\eqn\inv{\CI_{[\CF]} (X) = \Tr (-1)^F}
The right-hand side of this definition is the supersymmetric index in
the two-dimensional effective theory with the dynamical fields $\nu_k$
and the Calabi-Yau moduli (fields $\{ \phi_i \}$ and $\{ \sigma_j \}$)
being fixed.
The latter must satisfy the supersymmetry conditions \minimum\ in order
for $\CI_{[\CF]} (X)$ to be non-zero.
Since we regard $\phi_i$ and $\sigma_j$ as background fields,
the supersymmetric index (which in simple cases
counts the number of supersymmetric ground states in the
effective theory) is expected to be stable under deformations
of the four-fold $X$ \Windex.
Therefore, $\CI_{[\CF]} (X)$ must be a topological invariant of $X$,
at least as long as we do not cross a surface of marginal
stability\foot{The meaning of this term will become clear in
a moment when we give a `dual' definition of the invariants $\CI (X)$
in terms of supersymmetric cycles in $X$.} in the moduli space of $X$.
For a given $X$ and $[\CF] \in H^* (X, \Z)$, let $\CM_{[\CF]} (X)$ be
the set of solutions to \minv.
Then $\CI_{[\CF]} (X)$ is equal to the Euler number of $\CM_{[\CF]} (X)$:
\eqn\cinv{\CI_{[\CF]} (X) = \chi \Big( \CM_{[\CF]} (X) \Big)}

There is an equivalent geometric definition of $\CI$ in terms
of the homology class $[S]$ Poincar\'e dual to $[\CF]$.
With every supersymmetric vacuum representing the Ramond-Ramond
flux $\CF$ we can associate a BPS soliton constructed from a D-brane
\foot{Due to the anomaly inflow \GHM, D-branes on curved
manifolds carry induced charge of lower dimensional branes.
Therefore, the accurate relation between $\CI_{[\CF]} (X)$ and the invariant
$\CI_{[S]} (X)$ defined below involves {\it bound states} of D-branes.
This subtlety, however, is not important for the definition of
$\CI_{[S]} (X)$ unless we are interested in the precise
relation to $\CI_{[\CF]} (X)$.}
wrapped over the dual supersymmetric cycle $S$.
Hence, $\CI$ can be defined as the ``number" of supersymmetric
cycles with the homology class $[S]$.
More precisely, a system of invariants $\CI$ for a Calabi-Yau four-fold
$X$ is a family of maps $\CI \colon H_{*} (X, \Z) \to \Z$ defined by:
\eqn\sinv{\CI_{[S]} (X) = \chi \Big( \CM_{[S]} (X) \Big)}
where $\CM_{[S]} (X)$ is the moduli space of calibrated submanifolds
together with flat line bundles. It is important to bear in mind
that a D-brane carries a $U(1)$ gauge field.
We do not discuss here the important questions of smoothness
and compactness, simply assuming that there is a compactification
of $\CM_{[S]} (X)$ such that \sinv\ makes sense.
The motivation for this assumption is based in part on the following
two examples where a suitable compactification can be achieved.

$\underline{{\rm When}~ [S]~ {\rm is~ an~ element~ of}~ H_2 (X, \Z)}$,
the supersymmetric
cycles are given by the submanifolds calibrated with respect
to the K\"ahler form of $X$.
Therefore, in this case the invariant $\CI_{[S]} (X)$ is equal
to the Euler number of the moduli space of holomorphic curves
with a flat line bundle.
Unlike the definition of the Gromov-Witten invariants
\refs{\Ginv,\Winv}, there are no constraints on $S$ except to
stay within a given homology class.
Therefore one might worry that there
are infinitely many contributions from curves of arbitrary topology.
However, by the Riemann-Roch theorem the virtual (complex) dimension
of the moduli space of genus $g$ curves in a Calabi-Yau four-fold
is equal to $(1-g)$, so that generic point in $\CM_{[S]} (X)$
corresponds to a rational homology 2-sphere.
For such curves dimension of the space of flat $U(1)$ connections is zero.
At least for some simple cases, one may hope to use the close relation
to the Gromov-Witten invariants to compute $\CI_{[S]} (X)$.

Although it is difficult to count supersymmetric cycles in general case,
we can predict the asymptotic growth of $\CI_{[S]} (X)$ under
rescaling $[S] \to N [S]$ by a large number $N$.
To estimate the ``number" of holomorphic curves in a Calabi-Yau four-fold
$X$ we use the microscopic interpretation of black hole entropy \SV.
The analysis is very similar to the derivation of the entropy of extreme
black holes constructed from Calabi-Yau three-folds \refs{\MSW, \Ventropy}.
Consider M-theory compactification on a four-fold $X$
down to three dimensions.
Then, an M2-brane wrapped around a 2-cycle in the homology class
$[S] \in H_2 (X, \Z)$ represents the extreme BTZ black hole \BTZ\
with charge $[S]$. The entropy of this black hole is related
to the number of BPS states with a given charge:
\eqn\ient{S_{BTZ} = \log \CI_{[S]} (X)}
On the other hand, Bekenstein-Hawking entropy of the extreme
BTZ black hole is related to its mass $M_{BTZ}$ \Carlip:
\eqn\ment{S_{BTZ} \sim \sqrt{M_{BTZ}}}
In M theory the mass $M_{BTZ}$ is given by the mass of
a membrane wrapped around a cycle $S$ of minimal area,
so that as $[S] \to N  [S]$ the mass scales as $M_{BTZ} \sim N$.
Therefore, from \ment\ we conclude  that to the leading
order in $N \gg 1$ the number of BPS states grows as:
\eqn\ilarge{\CI_{[S]} (X) \sim \exp (a \sqrt{N})}
for some constant $a$.
There is one remark in place here.
The above argument based on the black hole entropy formula
actually predicts the growth of cohomologies of $\CM_{[S]} (X)$,
rather than its Euler number $\chi (\CM_{[S]} (X))$.
The latter should have the same large $N$ behavior unless there is
a nearly perfect cancellation between cohomologies of odd and even degree.
This is indeed the case for Calabi-Yau three-folds \Ventropy.
If such a miracle also happens for Calabi-Yau four-folds,
then the formula \ilarge\ should be understood only as an upper bound.
It says that $\CI_{[S]} (X)$ grow not faster than $\exp (a \sqrt{N})$.

$\underline{S~ {\rm is~ a~ special~ Lagrangian~ torus}}:$
Consider $[S] \in H_4 (X, \Z)$, so that $\CI_{[S]} (X)$
counts special Lagrangian submanifolds in the homology class $[S]$
with a choice of flat $U(1)$ connection.
A similar invariant counting special Lagrangian homology
3-spheres in Calabi-Yau three-folds was recently introduced by Joyce \J.
According to \refs{\Khom,\Vafa}, the mirror symmetry can be understood as
an equivalence between derived categories over the mirror spaces
$X$ and $\tilde X$.
In particular, it means that the moduli space of special Lagrangian
manifolds (with a flat connection) in a given homology class $[S] \in H_4(X)$
is equivalent to the moduli space, $\CM (V_{ch}, \tilde X)$,
of stable vector bundles over the mirror space $\tilde X$,
where the Chern class $ch \in H^*(\tilde X)$ is determined by $[S]$.
Therefore, when $S$ is a special Lagrangian manifold, eq. \sinv\
can be written as:
\eqn\vinv{\CI_{[S]} (X) = \chi \Big( \CM (V_{ch}, \tilde X) \Big)}

Now suppose that $X$ admits a special Lagrangian 4-torus fibration
\refs{\SYZ,\Gross}, and let us take $[S]$ to be the class of
the fiber. Then, according to the conjecture of Strominger,
Yau and Zaslow \SYZ, there is a compactification of $\CM_{[S]} (X)$
isomorphic to the mirror Calabi-Yau manifold:
$$
\CM_{[S]} (X) \cong \tilde X.
$$
In this case we have $\CI_{[S]} (X) = \chi (\tilde X) = \chi (X)$.
Obviously, the latter is a topological invariant of $X$, though not a new one.

\newsec{New Supersymmetric Vacua}

In the previous sections we found a set of constraints on Calabi-Yau
moduli and Ramond-Ramond fluxes which preserve $\CN=2$ supersymmetry
in two dimensions. There is a compact way of writing these conditions
in terms of the effective superpotentials \wone\ and \wtwo.

{}From the above discussion it is also clear that these supersymmetry
constraints are highly restrictive, so that usually it is very
difficult to find an explicit solution.
For example, when $G$ is the only non-zero field one has to look
for integral primitive forms of Hodge type $(2,2)$.
This problem was studied in a recent work \DRS.

Below we present a family of solutions to the supersymmetry
conditions with non-zero fluxes of $M$, $G$, and $\check F^{(8)}$.
For the sake of simplicity, we assume that the two-form flux $F$ and
the six-form flux $\check G^{(6)}$ vanish, so that superpotential is real.

Another insight from the supersymmetry condition \sctwo\
is that different components of Ramond-Ramond fields
which transform differently under the action of the $SU(4)$
holonomy group must vanish separately.
Indeed, by definition, a wedge product with the K\"ahler form $\CK$
takes one Ramond-Ramond form to another form (of degree greater
by 2) from the same irreducible representation.
Since the 0-form $M$ and the 8-form
$\check F^{(8)}$ are singlets under $SU(4)$, their contribution
in the effective superpotential can be canceled also only by
singlet component of the $G$-flux.
Such fields can be written as wedge products of the K\"ahler form $\CK$:
\eqn\vacone{ {M \over 2\pi} = n_0, \quad
{G \over 2\pi} = {n_4 \over 2} \CK \wedge \CK, \quad
{\check F^{(8)} \over 2\pi} =
{n_8 \over 4!} \CK \wedge \CK \wedge \CK \wedge \CK }

Of course, to consort with the flux quantization of the field $F$
one has to make an appropriate choice of the K\"ahler structure,
such that $\CK \in H^2(X, \Z)$ and $n_8$ is an integer number.
The quantization condition of the field $G$ is more subtle \Wflux:
$$
{G \over 2\pi} - {p_1(X) \over 4} \in H^2(X, \Z)
$$
When $\chi (X) /24 \in \Z$, the number $n_4$ is integer.
In massive Type IIA string theory
${M \over 2 \pi} = n_0$ also has to be an integer.

The ansatz \vacone\ automatically satisfies
the supersymmetry condition \scone\ in flat two-dimensional space.
Therefore, it remains to check the conditions that follow from
the twisted chiral superpotential \wtwo:
\eqn\condone{\int_X {M \over 4!} \CK \wedge \CK \wedge \CK \wedge \CK +
{1\over 2!} G \wedge \CK \wedge \CK + {\check F^{(8)}} = 0 }
and
\eqn\condtwo{{M \over 6} \CK \wedge \CK \wedge \CK + G \wedge \CK = 0}
Substituting the ansatz \vacone\ into \condone\ and \condtwo\
we obtain a system of two algebraic equations:
\eqn\algeq{n_0 + 6 n_4 + n_8 = 0, \quad {\rm and} \quad n_0  + 3 n_4 =0}
which must be solved in integer numbers. From the
second equation we see that there are no solutions when
$n_4 \in \Z + {1 \over 2}$. Hence, $\chi(X)$ has to be
divisible by 24, so that $n_4$ is an integer.
Then, the solution is given by:
\eqn\newvac{n_0 = n_8 = - 3 n_4.}
This solution can be easily generalized to include the Ramond-Ramond
two-form and six-form fluxes on the Calabi-Yau four-fold.

\appendix{I}{Massive Type IIA Supergravity on Eight-Manifolds}

In this appendix we present a microscopic test of our results.
In particular, we consider compactification of Type IIA string theory
on a Calabi-Yau four-fold $X$,
and ask for the conditions on the background Ramond-Ramond fields
to preserve at least ${\cal N}=2$ supersymmetry in two dimensions.
We follow the notations of \refs{\GVW,\BB} where a similar
analysis was carried out for compactifications with a $G$-field.

In the large volume limit the effective superstring dynamics is
described by a supergravity theory, {\it viz.} massive Type IIA
supergravity \Romans, since the 0-form field $M$ is generically non-zero.
The bosonic field content of the massive Type IIA supergravity
contains the metric $g_{MN}$, the dilaton $\varphi$, a vector field $A_M$,
tensor fields $B_{MN}$ and $C_{MNP}$ and a mass parameter $M$.
By supersymmetry we also have non-chiral spinor fields:
a gravitino $\psi_M$ and a dilatino $\lambda$.
The bosonic part of the Lagrangian looks like \Romans:
\eqn\alagr{ L = \sqrt{-g}\big[ - {1 \over 4} R +
{1 \over 2} (\nabla_M \varphi) (\nabla^M \varphi)
+ {1 \over 12} e^{- \varphi} H_{MNP} H^{MNP} +}
$$
+ {1 \over 48} e^{{1 \over 2} \varphi} G_{MNPQ} G^{MNPQ}
+ {1 \over 4} M^2 e^{{3 \over 2} \varphi} B_{MN} B^{MN} +
{1 \over 8} M^2 e^{{5 \over 2} \varphi} + \ldots \big]
$$
where we introduced the gauge-invariant field strengths \deight:
$$
F_{MN} = 2 \partial_{[M} A_{N]} + M B_{MN}
$$
$$
H_{MNP} = 3 \partial_{[M} B_{NP]}
$$
$$
G_{MNPQ} = 4 \partial_{[M} C_{NPQ]} + 24 B_{[MN} \partial_P A_{Q]}
+ 6 M^2 B_{[MN} B_{PQ]}
$$

Note that our notations slightly differ from the notations in \Romans.
One reason for this is to simplify the comparison with the ordinary
Type IIA supergravity. In the limit $M \to 0$ the last two
terms in  \alagr\ disappear and we end up with the standard
effective action for the massless Type IIA fields.
When $M \ne 0$, we can gauge away the vector field $A_M$
by gauge transformations of $B_{MN}$, leaving the latter
with a mass but without gauge invariance in a Higgs-type mechanism.
The value of the ten-dimensional cosmological constant in
the massive phase is also determined by $M$.
We use the metric in the Einstein frame which is related to
the string metric by the rescaling:
\eqn\frame{g^{\rm st}_{MN} = e^{ {1 \over 2} \varphi} g_{MN}}

The action of the massive Type IIA supergravity is invariant
under 16 left and 16 right supersymmetry transformations, such
that the left supersymmetries are chiral while the right supersymmetries
are anti-chiral, with the ten-dimensional chirality operator $\Gamma_{11}$.
Below we use the explicit form of the supersymmetry
transformations only for the fermionic fields.
For the gravitino we have:
\eqn\gravsusy{ \delta \psi_M = \nabla_M \eta -
{M \over 32} e^{{5 \over 4} \varphi} \Gamma_M \eta -
{1 \over 32} e^{{3 \over 4} \varphi} F_{NP}
({\Gamma_M}^{NP} - 14 \delta_M^N \Gamma^P) \Gamma_{11} \eta + }
$$
+ {1 \over 48} e^{ - {1 \over 2} \varphi} H_{NPQ}
({\Gamma_M}^{NPQ} - 9 \delta_M^N \Gamma^{PQ}) \Gamma_{11} \eta +
{1 \over 128} e^{ {1 \over 4} \varphi} G_{NPQR}
({\Gamma_M}^{NPQR} - {20 \over 3} \delta_M^N \Gamma^{PQR}) \eta
$$
and for the dilatino:
\eqn\dilsusy{ \delta \lambda =
- {1 \over 2 \sqrt{2}} (\partial_M \varphi) \Gamma^M \eta -
{5 \over 8 \sqrt{2}} M e^{{5 \over 4} \varphi} \eta +
{3 \over 8 \sqrt{2}} e^{{3 \over 4} \varphi} F_{MN} \Gamma^{MN}
\Gamma_{11} \eta + }
$$
+ {1 \over 12 \sqrt{2}} e^{ - {1 \over 2} \varphi} H_{MNP}
\Gamma^{MNP} \Gamma_{11} \eta -
{1 \over 96 \sqrt{2}} e^{ {1 \over 4} \varphi} G_{MNPQ}
\Gamma^{MNPQ} \eta
$$

In order to cancel the anomaly \sethrel\ one has to introduce
$N = {1 \over 24} \chi(X)$ fundamental strings filling
two-dimensional non-compact space.
We represent these strings by the following maximally
symmetric ansatz for the field strength $H_{MNP}$:
\eqn\hfield{H_{\mu \nu m} = \epsilon_{\mu \nu} \partial_m f(x^m)}
where the function $f(x^m)$ depends only on the coordinates on $X$.
A space-filling fundamental string is invariant under the
supersymmetry transformations \gravsusy\ - \dilsusy\ with the
supersymmetry parameter $\eta$ satisfying the projection relation:
\eqn\strings{\Gamma_{01} \Gamma_{11} \eta = \eta}
where we used the standard notation
$\Gamma_{M_1 \ldots M_n} = {1 \over n!} \Gamma_{[M_1} \ldots \Gamma_{M_n]}$
for the antisymmetrized product of gamma-matrices.
With this choice of sign, the relation \strings\ implies that
the supersymmetry is preserved by the spinor $\eta$ which has
positive eight-dimensional chirality.
In other words, if we decompose $\eta$ as:
\eqn\decomp{\eta = \epsilon \otimes \xi + \epsilon^* \otimes \xi^*,}
then the eight-dimensional commuting spinor $\xi$ must satisfy
$\gamma_9 \xi = \xi$,
where we also make the 10=2+8 split of the ten-dimensional gamma-matrices:
$$
\Gamma_{\mu} = \gamma_{\mu} \otimes \gamma_9, \quad
\Gamma_m = 1 \otimes \gamma_m
$$

The opposite choice of sign in the formula \strings\ would correspond to
anti-strings which make a negative contribution to the total charge $N$.
Since Calabi-Yau four-folds usually admit only one sort of
nowhere-vanishing chiral spinors\foot{Calabi-Yau four-folds that
admit spinors of both chiralities have zero Euler number.
According to the formula \sethrel, no new interesting vacua with
non-zero $G$-flux can be found in compactification on such $X$.},
the total number of strings, $N$, must be positive.
We choose $\xi$ to be a covariantly constant complex spinor of unit norm.
Note, the existence of such a spinor on the Calabi-Yau four-fold $X$
automatically implies $\CN=(2,2)$ supersymmetry in two dimensions.
Using the spinor $\xi$ we define the complex structure
${J_m}^n = i \xi^{\dagger} {\gamma_m}^n \xi$
and the K\"ahler form $\CK_{a \bar b} = i g_{a \bar b}$.
Since the metric on $X$ is of type (1,1), it is convenient to think
of `holomorphic' gamma-matrices $\gamma^a$ and $\gamma_a$
as creation and annihilation operators:
\eqn\annihil{ \gamma^a \xi =0, \quad \gamma_{a} \xi^* =0,
\quad \gamma^{\bar a} \xi^* =0, \quad \gamma_{\bar a} \xi =0}

The field $H$ is taken in the form \hfield, while all
the fermionic fields are assumed to vanish in the background.
As we will see in a moment,
a maximally symmetric compactification on $X$
with nontrivial Ramond-Ramond fields
typically leads to warped metric:
\eqn\ametric{ds_{10}^2 = \Delta^{-1} \Big( ds_2^2(x^{\mu}) +
ds_8^2 (x^m) \Big) }
where we introduced the warp factor $\Delta (x^m)$.
For now, both $\Delta(x^m)$ and $f(x^m)$ are scalar
functions of the coordinates on $X$. Below we show that
these two functions are related by the supersymmetry conditions.
Since two-dimensional space-time is assumed to be
maximally symmetric ({\it i.e.} flat Minkowski space-time,
de Sitter space, or anti-de Sitter space) the $\mu$-component
of the covariant derivative satisfies the following commutation
relation \GGWappear:
\eqn\ddcommlam{ [\nabla_{\mu}, \nabla_{\nu}] = \Lambda \CX_{\mu \nu}}
where $\Lambda$ is two-dimensional cosmological constant
and $\CX_{\mu \nu}$ is the Lorentz generator.

Using the proper rescaling of Dirac gamma-matrices:
$$
\matrix{ \Gamma_M & \to & \Delta^{-1/2} \Gamma_M, \quad
\Gamma^M & \to & \Delta^{1/2} \Gamma^M }
$$
we rewrite the supersymmetry variations \gravsusy-\dilsusy\ of the
fermionic fields in the metric \ametric. For the gravitino we get:
\eqn\gravsusy{ \eqalign{ \delta \psi_M = \nabla_M \eta &-
{1 \over 4} \partial_N(\log \Delta) {\Gamma_M}^N \eta
+ {\Delta \over 48} e^{ - {1 \over 2} \varphi}
({\Gamma_M} {\slash\!\!\!\! H} - 12 H_{MPQ} \Gamma^{PQ}) \Gamma_{11} \eta - \cr
&- {M \over 32} \Delta^{-1/2} e^{{5 \over 4} \varphi} \Gamma_M \eta -
{\Delta^{1/2} \over 32} e^{{3 \over 4} \varphi}
({\Gamma_M} {\slash\!\!\!\! F} - 16 F_{MP} \Gamma^P) \Gamma_{11} \eta + \cr
&+ {\Delta^{3/2} \over 128} e^{ {1 \over 4} \varphi}
({\Gamma_M} {\slash\!\!\!\! G} - {32 \over 3} G_{MPQR} \Gamma^{PQR}) \eta } }
%
And the variation of the dilatino in the metric \ametric\ looks like:
\eqn\dilsusy{ \eqalign{
\delta \lambda =
&- {\Delta^{1/2} \over 2 \sqrt{2}} (\partial_M \varphi) \Gamma^M \eta -
{5 \over 8 \sqrt{2}} M e^{{5 \over 4} \varphi} \eta +
{3 \over 8 \sqrt{2}} \Delta e^{{3 \over 4} \varphi} {\slash\!\!\!\! F}
\Gamma_{11} \eta + \cr
&+ {\Delta^{3/2} \over 12 \sqrt{2}} e^{ - {1 \over 2} \varphi}
{\slash\!\!\!\! H} \Gamma_{11} \eta -
{\Delta^2 \over 96 \sqrt{2}} e^{ {1 \over 4} \varphi}
{\slash\!\!\!\! G} \eta } }
where we use a short notation ${\slash\!\!\!\! G}$
for the total contraction $G_{MNPQ} \Gamma^{MNPQ}$, {\it etc.}

Now we put the variations of the fermion fields to zero and
project the corresponding equations onto subspaces of positive
and negative chirality. Let us assume that the spinor $\epsilon$
has positive chirality. The spinor $\epsilon$ of negative chirality
leads to the complex conjugated supersymmetry conditions.

We also have to make a decomposition \decomp\ of the spinor $\eta$ and
assume that the two-dimensional Killing spinor $\epsilon$ satisfies
an equation like
$\nabla_{\mu} \epsilon = \tilde m_{\psi} \gamma_{\mu} \epsilon
+ m_{\psi} \gamma_{\mu} \epsilon^*$,
where $m_{\psi}$ (resp. $\tilde m_{\psi}$) represents the mass
(resp. the twisted mass) of the two-dimensional gravitino field. From the
commutation relations \ddcommlam\
for the $\nabla_{\mu}$ it follows that either
$m_{\psi}$ or $\tilde m_{\psi}$ must be equal to zero.
Hence, one can distinguish two cases:

$i)$
The cosmological constant is given by
$\Lambda = - \vert \tilde m_{\psi} \vert^2$
and the Killing spinor $\epsilon$ satisfies:
\eqn\epstwisted{ \nabla_{\mu} \epsilon = \tilde m_{\psi} \gamma_{\mu} \epsilon}

$ii)$
The cosmological constant is given by
$\Lambda = - \vert m_{\psi} \vert^2$
and the Killing spinor $\epsilon$ satisfies:
\eqn\eps{ \nabla_{\mu} \epsilon = m_{\psi} \gamma_{\mu} \epsilon^*}

The second possibility corresponds to compactification
with non-zero value of the $(4,0)$ part of the background $G$-flux.
This follows from the $\mu$-component of \gravsusy\ which in this case
gives a relation between the warp factor $\Delta$ and the function $f$:
\eqn\dpsione{ (\partial_m \Delta) +
{1 \over 54} \Delta^2 e^{-{1 \over 2} \varphi} (\partial_m f) =0}
and:
\eqn\dpsitwo{ m_{\psi} \zeta^*
- {M \over 32} \Delta^{-1/2} e^{{5 \over 4} \varphi} \zeta
- {\Delta^{1/2} \over 16} e^{{3 \over 4} \varphi} {\slash\!\!\!\! F} \zeta
+ {\Delta^{3/2} \over 128} e^{ {1 \over 4} \varphi} {\slash\!\!\!\! G} \zeta =0}
Since the complex spinor $\zeta$ on a Calabi-Yau four-fold
obeys the following relation~\BBMOOY:
$$
\gamma^{abcd} \zeta = \Omega^{abcd} \zeta^*
$$
{}from the second equation \dpsitwo\ we get:
\eqn\ggholo{m_{\psi} =
{\Delta^{3/2} \over 128} e^{ {1 \over 4} \varphi} G_{abcd} \Omega^{abcd} }
Integrating over the entire Calabi-Yau four-fold $X$ and using
the self-duality of $G^{(0,4)}$, we obtain the formula for the gravitino mass:
\eqn\iiagravmass{m_{\psi} = {1 \over l_2} \int_X G \wedge \Omega }
where we introduced two-dimensional scale parameter
$l_2 = 128 \int d^8 x \Delta^{- 3/2} e^{ - {1 \over 4} \varphi}$.
Notice the striking similarity between \iiagravmass\ and the formula
for the effective superpotential $W \sim \int_X G \wedge \Omega$.
This is a non-trivial check of the proposed expression \wone\
for the effective superpotential induced by background $G$-flux.
Using microscopic arguments, here we found that compactifications
with $G^{(0,4)} \neq 0$ lead to two-dimensional vacua where
mass for the gravitino field is given by \iiagravmass.
In a supersymmetric situation masses of the bosonic modes of
the supergravity multiplet should be related to $m_{\psi}$ in
a supersymmetric fashion. In particular, one should expect a non-zero
cosmological constant $\Lambda = - \vert m_{\psi} \vert^2$.
However, it turns out that compactification with a $G_{4,0}$
flux alone leads to a solution with zero cosmological
constant and, therefore, implies broken supersymmetry \GVW.


However, Type IIA vacua with non-zero values of $m_{\psi}$
or $\tilde m_{\psi}$ have non-constant dilaton.
In fact, multiplying \gravsusy\ by $\Gamma^M$ from the left
and taking the trace over index `$M$' we obtain the following
supersymmetry condition:
\eqn\gravtr{ \eqalign{
& \Gamma^M \nabla_M \eta -
{9 \over 4} {\slash\!\!\! \partial} (\log \Delta) \eta
- {\Delta \over 24} e^{ - {1 \over 2} \varphi} {\slash\!\!\!\! H} \Gamma_{11} \eta - \cr
& - {5 \over 16} M \Delta^{-1/2} e^{{5 \over 4} \varphi} \Gamma_M \eta
+ {3 \over 16} \Delta^{1/2} e^{{3 \over 4} \varphi} {\slash\!\!\!\! F} \Gamma_{11} \eta
+ {1 \over 192} \Delta^{3/2} e^{ {1 \over 4} \varphi} {\slash\!\!\!\! G} \eta =0 } }
The second line of this equation looks very similar to the variation
of the dilatino \dilsusy. In fact, from the linear combination of \gravtr\
and \dilsusy\ we get:
\eqn\dildirac{\Gamma^M \nabla_M \eta = - {1 \over 4}
({\slash\!\!\! \partial} \varphi) \eta }
Here, for the sake of simplicity, we assumed that $\Delta$ and $f$
are constant\foot{Strictly speaking, $\Delta$ and $f$ can not be
constant in a generic compactification on a Calabi-Yau space $X$.
Indeed, the warp factor has to satisfy Laplace equation
in eight compact dimensions. However, integrating over a compact
manifold $X$, the terms containing derivatives of these functions
with respect to internal coordinates average to zero.}.
For example, in the gauge we are using, where
the Killing spinor $\epsilon$ satisfies \epstwisted, we obtain:
\eqn\ddeq{(8 \tilde m_{\psi} + {\slash\!\!\! \partial} \varphi) \epsilon = 0}
This equation implies that:
$$
(\partial \varphi)^2 = 64 \vert m_{\psi} \vert^2
$$
In the gauge we are using a solution to these equations
is given by a linear dilaton vacuum \BGV:
$$
\varphi = 8 \tilde m_{\psi} x^1
$$
where the constant spinor $\epsilon$ is supposed to satisfy:
$$
(\gamma^1 + 1)\epsilon = 0
$$
This dilaton background spontaneously breaks
Poincare invariance in two dimensions (down to time translations)
and half the supersymmetries preserved by Calabi-Yau four-fold.
The physics of these vacua is very interesting and quite subtle \BGV.
Here we shall not discuss it any further, and return to the test
of supersymmetry conditions for the new solutions found in section 5.

As in the solutions \newvac, we assume that
gravitino fields are massless. This implies that dilaton depends
only on the internal coordinates $x^m$, but not on
the two-dimensional coordinates $x^{\mu}$, {\it cf.} \dildirac.
This dependence can be determined from eq. \dpsione\ and from
the variation of the dilatino \dilsusy\ that gives:
\eqn\dlone{ (\partial_m \varphi) -
{\Delta \over 6} e^{ - {1 \over 2} \varphi} (\partial_{m} f) =0}
The solution looks like:
\eqn\fsol{e^{\varphi} = \Delta^{ - 9}, \quad
f = {108 \over 11} \Delta^{ - 11/2}}
As we discussed above, the other components of \dilsusy\ are
equivalent to the linear combination of \dpsitwo\ and \gravtr,
so that we may focus only on the latter two conditions
and no longer worry about the variation of the dilatino --
it will vanish automatically, as long as the scalar functions
$f(x^m)$, $\Delta (x^m)$, and $\varphi (x^m)$ obey \fsol.
In what follows we assume that this is always the case,
and suppress the dependence on these functions in order
to make a comparison to the result of the previous subsection,
where we implicitly set the string coupling constant,
$g_{\rm st} = e^{\varphi}$, equal to unity.

The remaining components of the supersymmetry
conditions \dpsitwo\ and \gravtr\ depend on the particular
choice of Ramond-Ramond fluxes.
We shall analyze these conditions for the background fluxes given
by the wedge products of the K\"ahler form $\CK$ as in \vacone.
Then, the only non-trivial components of \dpsitwo\ and \gravtr\
correspond to singlet representation of $SU(4)$ and have a very
simple form:
\eqn\susycone{\Big( - 4 M +
28 F_{\mu \nu} \gamma^{\mu \nu} \Gamma_{11}
+  {\slash\!\!\!\! G} \Big) \eta =0}
and
\eqn\susyctwo{\Big(12 M +
12 F_{\mu \nu} \gamma^{\mu \nu} \Gamma_{11}
+  {\slash\!\!\!\! G} \Big) \eta =0}

Notice, that with our choice of the two-form flux $F$ allows to
rewrite these supersymmetry conditions without $\Gamma_{11}$,
which is related to the fact that $\check F^{(8)}$ has a real
contribution to the effective superpotential \wtwo.
In fact, using the standard gamma-matrix algebra and
the projection relation \strings\ we get the following relation:
\eqn\fhatfhat{F_{\mu \nu} \gamma^{\mu \nu} \Gamma_{11} \eta = n_8 \eta}
where we identified $F_{\mu \nu}$ components of the 2-form field
strength with the Hodge dual 8-form flux $\check F^{(8)}$ in \vacone.
Using the relation \fhatfhat\ and the following properties of $SU(4)$
holonomy manifolds \BBMOOY:
\eqn\sufouridnt{
\gamma_{\bar a \bar b c d} \zeta = 3 \CK_{[ \bar a c} \CK_{\bar b d]} \zeta,
\quad ^* (\CK \wedge \CK) = \CK \wedge \CK,}
one can put \susycone\ - \susyctwo\ in the form of two
algebraic equations:
$$
- 4 n_0 + 28 n_8 + 72 n_4 = 0
\quad {\rm and} \quad
12 n_0 + 12 n_8 + 72 n_4 = 0
$$

Now it is easy to solve these equations:
$$
n_0 = n_8 = -3 n_4
$$
This solution is equivalent to \newvac\ that we found
earlier analyzing superpotential terms in the effective two-dimensional theory.
Therefore, microscopic analysis of the supersymmetry conditions
in massive Type IIA supergravity gives us an independent evidence
for the expressions \wone\ and \wtwo.

Finally, we explain that another set of supersymmetry transformations
\gravsusy\ and \dilsusy\ with a spinor $\eta$ of negative chirality
leads to complex conjugated supersymmetry conditions.
Notice that according to \strings\ the ten-dimensional chirality
of the spinor $\eta$ is directly related to the two-dimensional
chirality of the spinor $\epsilon$.
If in our calculation we used the spinor $\epsilon$ of negative chirality,
we would obtain the complex conjugated supersymmetry conditions.
To see this we note that $\Gamma_{11}$ appears in \gravsusy\ and
\dilsusy\ only in the terms that contain $F_{MN}$ or $H_{MNP}$.
However, the three-form $H$ always comes contracted with two-dimensional
gamma-matrices, so that we have extra $\gamma_{3}$ factor due to
the specific form of the ansatz \hfield.
Therefore, the terms with $H$ do not change the sign if we
change the chirality of $\eta$ (= chirality of $\epsilon$).
This is not the case for the terms proportional to the internal components
of the two-form $F$ and for the terms linear in $G_{\mu \nu mn}$
which have the same gamma-matrix structure:
\eqn\ghatfhat{G_{\mu \nu m n} \Gamma^{\mu \nu m n} \eta =
- G_{01 m n} \gamma^{m n} \Gamma_{11} \eta}
We conclude that the supersymmetry conditions for the spinors
of negative chirality differ only by signs of all the terms with
$F_{mn}$ and $G_{\mu \nu mn}$. These conditions should be compared
with the variation of complex conjugate superpotential \wtwo.
In our examples of the form \vacone, these conditions are
automatically satisfied because the fields $F_{mn}$ and
$G_{\mu \nu mn}$ were assumed to vanish in the background.


\vskip 30pt
\centerline{\bf Acknowledgments}

I would like to express my gratitude to
S.~James Gates and especially to Edward Witten
for helpful discussions and constant encouragement.
The work was supported in part by the Caltech Discovery Fund,
grant RFBR No 98-02-16575 and
Russian President's grant No 96-15-96939.

\listrefs
\end